\documentclass[a4paper,UKenglish]{lipics}
\usepackage{amsmath}
\usepackage{amssymb}
\usepackage{amsfonts}
\usepackage{bussproofs}
\graphicspath{{./graphics/}}

\newcounter{c}
\newcommand{\nentry}[1]{\noindent\textbf{#1}}

\newcommand{\note}[1]{\textsl{#1}}
\newcommand{\dit}{$\triangleright$}

\newcommand{\cnst}[1]{\textsl{#1}}
\newcommand{\grdef}{\ensuremath{::=\ }}
\newcommand{\lett}[1]{\texttt{#1}}

\newcommand{\sg}[1]{\textsf{#1}}
\newcommand{\formtype}{\ensuremath{A}}
\newcommand{\sftype}{\ensuremath{\Sigma}}
\newcommand{\bnot}{\ensuremath{!}}
\newcommand{\band}{\ensuremath{\&\&}}
\newcommand{\mean}[1]{\ensuremath{\hat{#1}}}
\newcommand{\op}[1]{\ensuremath{\overline{#1}}}
\newcommand{\interprt}{\ensuremath{\mathcal{I}}}
\newcommand{\vars}{\ensuremath{\mathcal{V}}}
\newcommand{\bool}{\ensuremath{\mathsf{Bool}}}
\newcommand{\vval}[2]{\ensuremath{[\![ #1]\!]^{#2}}}

\newcommand{\sat}[2]{\ensuremath{#1\models#2}}

\newcommand{\allmodels}{\ensuremath{\mathcal{M}}}
\newcommand{\pow}{\ensuremath{\mathcal{P}}}
\newcommand{\finpow}{\ensuremath{\pow_\mathsf{fin}}}
\newcommand{\iset}[1]{\ensuremath{\{i\in\interprt \mid #1\}}}
\newcommand{\ssplit}[2]{\ensuremath{#1\!\mid\!#2}}
\newcommand{\lmods}[1]{\ensuremath{\hat{#1}}}

\newcommand{\ruleq}[1]{\ensuremath{\mathsf{#1}}}
\newcommand{\ttabl}[1]{\ensuremath{\mathcal{T}({#1})}}

\newcommand{\Mwedge}{\ensuremath{{\wedge\hspace{-0.65em}\wedge}\hspace{-0.65em}{\wedge\hspace{-0.65em}\wedge}}}
\newcommand{\Mvee}{\ensuremath{{\vee\hspace{-0.65em}\vee}\hspace{-0.65em}{\vee\hspace{-0.65em}\vee}}}

\title{Presentation of Classical Propositional Tableaux on Program Design Premises\footnote{This work was partially supported by ANII--Agencia Nacional de Investigaci\'on e Innovaci\'on, Uruguay.}}
\author{Juan Michelini}
\author{\'Alvaro Tasistro}
\affil{Universidad ORT Uruguay\\  \texttt{michelini@ort.edu.uy}, \texttt{tasistro@ort.edu.uy}}
\authorrunning{J. Michelini and \'A. Tasistro}

\Copyright{Juan Michelini and \'Alvaro Tasistro}

\subjclass{}
\keywords{}

\serieslogo{logo_ttl}
\volumeinfo
  {M. Antonia {Huertas}, Jo\~ao {Marcos}, Mar\'ia {Manzano}, Sophie {Pinchinat}, \\
  Fran\c{c}ois {Schwarzentruber}}
  {5}
  {4th International Conference on Tools for Teaching Logic}
  {1}
  {1}
  {137}
\EventShortName{TTL2015}

\begin{document}
\maketitle
\begin{abstract}
We propose a presentation of classical propositional tableaux elaborated by application of methods that are noteworthy in program design, namely program derivation with separation of concerns.  We start by deriving from a straightforward specification an algorithm given as a set of recursive equations for computing all models of a finite set of formulae. 
Thereafter we discuss the employment of data structures, mainly with regard to an easily traceable manual execution of the algorithm. This leads to the kinds of trees given usually as constituting the tableaux.
The whole development strives at avoiding gaps, both of logical and motivational nature.
\end{abstract}

\section{Introduction}

We teach a course \textit{Logic for Computing} in a Software Engineering programme of studies. Prior to this, students have received courses in Calculus, Algebra and introductory Programming in Java, plus a course called \textit{Foundations of Computing}
, which introduces  polymorphic, higher-order functions and inductive types with the fundamental methods of  induction and recursion in their various forms. \textit{Foundations of Computing} makes emphasis on a mathematical approach to Programming, specifically on correctness proofs.
 \textit{Logic for Computing}, in turn, concerns itself essentially with the notion of \emph{formal} proof.

It follows from the foregoing that we should be very much interested in making explicit \emph{methods} of proof. By this we mean both general strategies for developing and fully understanding solutions to problems, as well as manners of presenting the corresponding proofs which convey natural, concise and complete justifications of their design.
Now, as it turns out, we have observed that some methods that have arisen within what could be called the science of Programming can be employed for obtaining or conveniently presenting mathematical results. This is to our mind a fact to be most welcome, for it exposes a unity of method between Programming and Mathematics that cannot but bring about positive outcomes for both sides, at least in as much the learning and teaching aspects are concerned.

In this paper we present an example of the latter, concerning the presentation of the method of \textit{tableaux}.
This is a proof procedure for both propositional and predicate logic dating back to \cite{Hintikka} and \cite{Beth}, and whose ultimate variant (termed \textit{analytic} tableaux) has been introduced in \cite{Smullyan}. 
Specifically, what we do is: 
(1) We derive the method as a set of equations ---to be used as rewriting rules--- from a straightforward specification, namely the one demanding the computation of the set of all models of the given set of formulas.
(2) We discuss the design of data structures for actually effecting and keeping trace of the execution of the method, which leads to the sorts of trees that are called ``the tableaux'' in textbooks.
The first part yields a compact proof of the correctness of the method, much simpler than the ones in textbooks.
The second part introduces the convenient and classical notation and establishes its correctness relating it to the set of equations originally given by a simple inductive argument. 
As a whole, the process is one in which we repeatedly employ simple techniques of \emph{program derivation} and \emph{separation of concerns} to obtain a presentation and justification both modular and simpler of the method of tableaux. 

The rest of the paper consists of a general background section whose contents is assumed to be taught priorly to the study of tableaux. In section 3 we present the derivation of the equational algorithm calculating the set of all models of given set of formula. In section 4 we discuss the data structures for tracing the execution of the algorithm, leading to the usual presentations of tableaux, after which we finish up with a general discussion. The presentation is to be read basically as a concise course handout, with some explicit considerations of logical or didactic nature. 
\section{Background} 

\nentry{Syntax.} It is enough to consider the set of connectives \{$\neg$ , $\wedge$\}. Then the set of formul\ae\ is defined as usual, starting out from a denumerable set \vars\ of propositional letters \lett{p}:

$\alpha , \beta \grdef \lett{p} \mid \neg\alpha \mid \alpha \wedge \beta.$

\noindent We use \textit{signed} formulae
$\sigma \grdef  S \alpha$ where $S \grdef \sg{F} \mid \sg{T}$, as the forms of assertion or judgement.\footnote{The use of signed formulae avoids privileging one boolean value over the other and consequently simplifies definitions.}

\nentry{Semantics.} 
Interpretations belong in 
\interprt $= \vars \rightarrow \bool$. The \textit{semantic value} of each formula is defined as follows ---
let \formtype\ be the set of formulae and (\bnot) and (\band) denote respectively Boolean negation and conjunction:

$\vval{\_}{\_} :: \formtype \rightarrow \interprt \rightarrow \bool$

$\begin{array}[b]{lll}
\vval{\lett{p}}{i} & = & i\,\lett{p} \\
\vval{\neg\alpha}{i} & = & \bnot\,\vval{\alpha}{i} \\
\vval{\alpha\wedge\beta}{i} & = & \vval{\alpha}{i}\ \band\ \vval{\beta}{i}
\end{array}$.

 Using the former we now define \textit{truth} of an assertion (signed formula) in an interpretation. 
Call \mean{S} the boolean value corresponding to sign $S$. Then
$\sat{i}{S\alpha} \equiv \vval{\alpha}{i} = \mean{S}$,
\noindent which reads: $i$ is a \textit{model} of $S\alpha$, and also: $i$ \textit{satisfies} $S\,\alpha$ or $S\,\alpha$ is \textit{valid in} $i$.
We shall consider \emph{finite} sets $\Gamma$ of signed formulae and define models thereof (i.e. \sat{i}{\Gamma}) as the interpretations satisfying every formula of $\Gamma$.

\nentry{Truth in an interpretation.} 
It is generally interesting to develop a method for checking truth of signed formulae in an interpretation.
If we start with the propositional letters, we get:

$\sat{i}{S\lett{p}} 
\equiv^{\mathrm{\ (model\ of\ signed\ formula)}}$

$\vval{\lett{p}}{i} = \mean{S}
\equiv^{\mathrm{\ (semantic\ function)}}$

i\,\lett{p} = \mean{S}.

For the other cases we wish to obtain (structurally) recursive equations.
As to negation, writing \op{S} the sign opposite to $S$, we obtain:

$\sat{i}{S(\neg\alpha)} 
\equiv^{\mathrm{\ (model\ of\ signed\ formula)}}$

$\vval{\neg\alpha}{i} = \mean{S}
\equiv^{\mathrm{\ (semantic\ function)}}$

$\bnot\,\vval{\alpha}{i} = \mean{S}
\equiv^{\mathrm{\ (negating\ both\ sides\ to\ isolate\ \vval{\alpha}{i})}}$

$\vval{\alpha}{i} = \ \bnot\,\mean{S}
\equiv^{\mathrm{\ (opposite\ sign)}}$

$\vval{\alpha}{i} = \mean{\op{S}}
\equiv^{\mathrm{\ (model\ of\ signed\ formula)}}$

\sat{i}{\op{S}\,\alpha}.

\noindent Finally, turning to conjunction:

$\sat{i}{S(\alpha\wedge\beta)} 
\equiv^{\mathrm{\ (model\ of\ signed\ formula)}}$

$\vval{\alpha\wedge\beta}{i} = \mean{S}
\equiv^{\mathrm{\ (semantic\ function)}}$

\vval{\alpha}{i}\ \band\ \vval{\beta}{i}  = \mean{S},

\noindent where we seem to get stuck. Indeed, to rewrite the left hand side requires to consider the definition of (\band) and this is not uniform with respect to truth and falsity.  
Therefore we are led to try instead distinguishing the cases of $S$:

\note{Case \sg{T}}:

\sat{i}{\sg{T}(\alpha\wedge\beta)} 
$\equiv^{\mathrm{\ (calculation\ above)}}$

\vval{\alpha}{i}\ \band\ \vval{\beta}{i}  = \cnst{True}
$\equiv^{\mathrm{\ (definition\ of\ \band)}}$

\vval{\alpha}{i}= \cnst{True} $\Mwedge$\footnote{We distinguish $\wedge$ from $\Mwedge$, the first being the conjunction in the object language, the second being the conjunction in the meta-language. Likewise with $\vee$ and $\Mvee$.} \vval{\beta}{i}  = \cnst{True}
$\equiv^{\mathrm{\ (satisfaction)}}$

\sat{i}{\sg{T}\,\alpha} $\Mwedge$ \sat{i}{\sg{T}\,\beta}.


\note{Case \sg{F}}:

\sat{i}{\sg{F}(\alpha\wedge\beta)} 
$\equiv^{\mathrm{\ (calculation\ above)}}$

\vval{\alpha}{i}\ \band\ \vval{\beta}{i}  = \cnst{False}
$\equiv^{\mathrm{\ (property\ of\ \band)}}$

\vval{\alpha}{i}= \cnst{False} $\Mvee$ \vval{\beta}{i}  = \cnst{False}
$\equiv^{\mathrm{\ (satisfaction)}}$

\sat{i}{\sg{F}\,\alpha} $\Mvee$ \sat{i}{\sg{F}\,\beta}.


\noindent Ultimately, we arrive at the following characterisation of the satisfaction relation:


$\begin{array}{llll}
\mathrm{Signed\ letter:} & \sat{i}{S\lett{p}} & \equiv & i\,\lett{p} = \mean{S}\\

%

\mathrm{Signed\ negation:} & \sat{i}{S\,(\neg\alpha)} & \equiv & \sat{i}{\op{S}\,\alpha} \\

\mathrm{True\ conjunction:} & \sat{i}{\sg{T}(\alpha\wedge\beta)} & \equiv & \sat{i}{\sg{T} \alpha}\ \Mwedge \ \sat{i}{\sg{T} \beta} \\

\mathrm{False\ conjunction:} & \sat{i}{\sg{F}(\alpha\wedge\beta)} & \equiv & \sat{i}{\sg{F} \alpha}\ \Mvee \ \sat{i}{\sg{F} \beta}.
\end{array}$


%
%
%
%

\section{The Set of All Models}
We now set ourselves the problem of computing \emph{all models} of any given finite set $\Gamma$ of signed formulae. This is accomplished by the function

$\allmodels :: \finpow({\sftype}) \rightarrow \pow({\interprt})$

$\allmodels (\Gamma) = \iset{\sat{i}{\Gamma}}$,

\noindent where \sftype\ is the set of signed formulae, \pow{} is the power set operator yielding the set of subsets of given set, and \finpow{} 
does the latter for the finite subsets\footnote{Our use of ``computing'' seems at first generous indeed, since we are setting ourselves to generating in general infinite sets of infinite objects. Consider for that matter the case $\Gamma = \{\lett{p}\}$. Then any interpretation assigning \cnst{True} to \lett{p} is a member of the answer set. We shall see later how to settle this issue in detail ---the general idea is to give a finite sufficient characterisation of infinite sets of interpretations.}.
Now this straightforward definition presents the inconvenience that, as a method of computation, it obliges to construct all the interpretations and check each of them against the formulae\ in $\Gamma$. We are rather in the search of a syntactic procedure, i.e. one that applied exclusively to the formulae in $\Gamma$ ends up arriving at the desired set of models. Let us then examine $\Gamma$.

First of all, $\Gamma$ could be empty, which is indeed a plainly uninteresting case. Indeed, every interpretation trivially satisfies the empty set of formulae and so the result in such case is \interprt. So let us assume $\Gamma\neq\emptyset$. If this is the case, then we can pick any one of the formulae $\sigma$ in $\Gamma$ and write the latter in the form \ssplit{\Delta}{\sigma}, which means that 
$\Gamma = \Delta \cup \sigma$ and 
$\sigma \not\in \Delta$. 
Given the former, we can now write:

$\allmodels(\Gamma)
=^{\mathrm{\ (split\ \Gamma)}}$

$\allmodels\,(\ssplit{\Delta}{\sigma})
=^{\mathrm{\ (definition\ of\ \allmodels)}}$

$\iset{\sat{i}{\ssplit{\Delta}{\sigma}}}
=^{\mathrm{\ (satisfaction\ of\ a\ set\ of\ formulae)}}$

\iset{\sat{i}{\Delta} \Mwedge \sat{i}{\sigma}}.


The only source of information in the latter expression is the analysis of the form of $\sigma$, and so we are led to an examination of cases, i.e. to considering:

$\allmodels\,(\ssplit{\Delta}{S\,\lett{p}})  = \iset{\sat{i}{\Delta} \Mwedge \sat{i}{S\,\lett{p}}}$,

$\allmodels\,(\ssplit{\Delta}{S\,(\neg\alpha)})  = \iset{\sat{i}{\Delta} \Mwedge \sat{i}{S\,(\neg\alpha)}}$,

$\allmodels\,(\ssplit{\Delta}{S\,(\alpha\wedge\beta)})  = \iset{\sat{i}{\Delta} \Mwedge \sat{i}{S\,(\alpha\wedge\beta)}}$.

\noindent We can make profit of this analysis by using the results obtained at the end of the preceding section for checking the truth of signed formulae in a given interpretation.
As it happens, the first case is a bit discouranging, for the satisfiability condition \sat{i}{S\,\lett{p}} takes us to consider the value of \lett{p} in the given interpretation, i.e. a semantic rather than a syntactic move. But it pays off to insist. Negation gives the following:


$\allmodels\,(\ssplit{\Delta}{S\,(\neg\alpha)})
{=}^{\mathrm{\ (inferred\ above)}}$

$\iset{\sat{i}{\Delta} \Mwedge \sat{i}{S\,(\neg\alpha)}}
{=}^{\mathrm{\ (satisfaction\ of\ signed\ negation)}}$

$\iset{\sat{i}{\Delta} \Mwedge \sat{i}{\op{S}\,\alpha}}
{=}^{\mathrm{\ (satisfaction\ of\ a\ set\ of\ formulae)}}$

$\iset{\sat{i}{\Delta, \op{S}\,\alpha}}
{=}^{\mathrm{\ (definition\ of\ } \allmodels)}$

$\allmodels\,(\Delta, \op{S}\,\alpha)$,


\noindent where we have used $(,)$ instead of $(\cup)$ for set union. 
Notice that it is indeed this operation and not the formerly used split $(\mid)$ which is to be employed in this case, for we do not now know whether the formula $\op{S}\,\alpha$
belongs or not to $\Delta$.
The equation thus obtained, namely

$\allmodels\,(\ssplit{\Delta}{S\,(\neg\alpha)}) = \allmodels\,(\Delta, \op{S}\,\alpha)$,


\noindent is very convenient, for it rewrites the desired set of all models into an expression in which the overall complexity of the formulae has been strictly decreased. The same works for conjunction, whose results with respect to satisfiability can be used by distinguishing the two cases of the sign affecting it:

%
%
%
%
%
%

$\allmodels\,(\ssplit{\Delta}{\sg{T}(\alpha\wedge\beta)})
=^{\mathrm{\ (inferred\ above)}}$

$\iset{\sat{i}{\Delta} \Mwedge \sat{i}{\sg{T}(\alpha\wedge\beta)}}
=^{\mathrm{\ (truth\ of\ conjunction)}}$

$\iset{\sat{i}{\Delta} \Mwedge \sat{i}{\sg{T}\,\alpha} \Mwedge \sat{i}{\sg{T}\,\beta}}
=^{\mathrm{\ (satisfaction\ of\ a\ set\ of\ formulae)}}$

$\iset{\sat{i}{\Delta, \sg{T}\,\alpha,\sg{T}\,\beta}}
=^{\mathrm{\ (definition\ of\ \allmodels)}}$

$\allmodels\,(\Delta, \sg{T}\,\alpha,\sg{T}\,\beta)$.

\noindent On the other hand:


$\allmodels\,(\ssplit{\Delta}{\sg{F}(\alpha\wedge\beta)})
=^{\mathrm{\ (inferred\ above)}}$

$\iset{\sat{i}{\Delta} \Mwedge \sat{i}{\sg{F}(\alpha\wedge\beta)}}
=^{\mathrm{\ (falsity\ of\ conjunction)}}$

$\iset{\sat{i}{\Delta} \Mwedge (\sat{i}{\sg{F}\,\alpha} \Mvee \sat{i}{\sg{F}\,\beta}})
=^{\mathrm{\ (distributing\ } \Mwedge\ \mathrm{ over\ } \Mvee)}$

$\iset{(\sat{i}{\Delta} \Mwedge \sat{i}{\sg{F}\,\alpha}) \Mvee (\sat{i}{\Delta} \Mwedge \sat{i}{\sg{F}\,\alpha})}
=^{\mathrm{\ (trading\ } \Mvee\ \mathrm{for\ } \cup\  \mathrm{out\ of\ set\ comprehension)}}$

$\iset{\sat{i}{\Delta} \Mwedge \sat{i}{\sg{F}\,\alpha}} \cup \iset{\sat{i}{\Delta} \Mwedge \sat{i}{\sg{F}\,\beta}}
=^{\mathrm{\ (satisfaction\ of\ a\ set\ of\ formulae)}}$

$\iset{\sat{i}{\Delta, \sg{F}\,\alpha}} \cup \iset{\sat{i}{\Delta, \sg{F}\,\beta}}
=^{\mathrm{\ (definition\ of\ \allmodels)}}$

$\allmodels\,(\Delta, \sg{F}\,\alpha) \cup \allmodels\,(\Delta, \sg{F}\,\beta)$.


\noindent As a result we have so far obtained:


$\allmodels\,(\ssplit{\Delta}{S\,(\neg\alpha)})  = \allmodels\,(\Delta, \op{S}\,\alpha)$

$\allmodels\,(\ssplit{\Delta}{\sg{T}(\alpha\wedge\beta)}) = \allmodels\,(\Delta, \sg{T}\,\alpha,\sg{T}\,\beta)$

$\allmodels\,(\ssplit{\Delta}{\sg{F}(\alpha\wedge\beta)})  = \allmodels\,(\Delta, \sg{F}\,\alpha) \cup \allmodels\,(\Delta, \sg{F}\,\beta)$,


\noindent where the case of a signed letter, i.e. a \emph{literal}, could not be included. Now taking a look at the preceding equations for \allmodels, we readily realize that the missing case is actually that of a set not containing any composite formulae, i.e. that of a set of literals. Such is the base case of our recursion, since this proceeds by decreasing the size of the formulae of the set being treated ---and not the size of the set itself.
Therefore it is natural to wonder whether the solution of such base case could actually be just immediate.
This is indeed the case, because there is a straightforward manner of converting a set $\Gamma$ of literals into the set of all its models. There are two cases:\begin{description}
\item[\dit] $\Gamma$ contains pairs of opposite literals. Then it is inconsistent and the set of its models is $\emptyset$.
\item[\dit] Otherwise the models of $\Gamma$ are the interpretations that coincide with $\Gamma$ at the letters mentioned in it.
\end{description}
Formally, call \lmods{\Gamma} the set of all models of the set $\Gamma$ of literals. It is defined as follows:

$\lmods{\Gamma} = \left\{ \begin{array}{ll}
	\emptyset & \mathrm{if}\ \{S\,\lett{p}, \op{S}\,\lett{p}\} \subseteq \Gamma\ \mathrm{for\ some}\ \lett{p}, \\
	\iset{(\forall S\,\lett{p} \in \Gamma)\ i\,\lett{p} = \mean{S}} & \mathrm{otherwise.}
\end{array}\right.$

\noindent Notice that the alternative is decidable and that in the second case the result is sufficiently characterised by the set $\Gamma$ of literals and so we get a finite representation of it.
We can then put together equations for actually computing \allmodels:


$\allmodels\,(\ssplit{\Delta}{\sg{T}(\alpha\wedge\beta)}) = \allmodels\,(\Delta, \sg{T}\,\alpha,\sg{T}\,\beta)$

$\allmodels\,(\ssplit{\Delta}{\sg{F}(\alpha\wedge\beta)})  = \allmodels\,(\Delta, \sg{F}\,\alpha) \cup \allmodels\,(\Delta, \sg{F}\,\beta)$

$\allmodels\,(\ssplit{\Delta}{S\,(\neg\alpha)})  = \allmodels\,(\Delta, \op{S}\,\alpha)$


$\allmodels(\Gamma) = \lmods{\Gamma}$ if $\Gamma$ is a set of literals.

We claim that \allmodels\ captures the essence of the method of tableaux, and the derivation carried out above gives actually a quite simple proof of its correctness. Nevertheless, its actual execution needs to employ some kind of data structure to record the successive transformations leading to the final result. That is what we turn now to examining.

\section{Data Structures for the Tableaux}
\nentry{List of lists.} If we ignored the second equation above we would be in presence of a tail-recursive algorithm,  i.e. one whose execution could consist merely in successively rewriting the finite set of formulae at hand. We would then do simply with a list of formulae from which we would choose the next formula to be transformed. Now, consideration of the second equation does not in principle introduce any dramatic modification of this situation: it is enough that each application of the equation produces a split of the list from which the formula $\sg{F}(\alpha\wedge\beta)$ is taken into two lists, each of it containing exactly one of the two formulae $\sg{F}\alpha$ and $\sg{F}\beta$ in place of the original one, without any further change.

We illustrate this  by means of an example. Suppose we consider
 $\Gamma = [\sg{T}(\lett{p}\wedge\neg\lett{q}) , \sg{F}(\lett{p}\wedge\lett{q})]$, 
\noindent which we already make into a \emph{list} of formulae (as indicated by the use of the $[\ldots]$ notation).

\noindent Then we may choose say the second formula to proceed, which leads us to employ \allmodels's second equation splitting the original list into two, yielding
$[\sg{T}(\lett{p}\wedge\neg\lett{q}) , \sg{F}\lett{p}] , [\sg{T}(\lett{p}\wedge\neg\lett{q}) , \sg{F}\lett{q}]$.\\
In the next step we may choose any one of the two occurrences of the only composite (i.e. non-literal) formula. Say  we take the left one. The equation to employ is \allmodels's first, yielding
$[\sg{T}\lett{p} , \sg{T}(\neg\lett{q}) , \sg{F}\lett{p}] , [\sg{T}(\lett{p}\wedge\neg\lett{q}) , \sg{F}\lett{q}]$.
Of course we may do the same with the right occurrence of the formula just considered, arriving at
$[\sg{T}\lett{p} , \sg{T}(\neg\lett{q}) , \sg{F}\lett{p}] , [\sg{T}\lett{p} , \sg{T}(\neg\lett{q}) , \sg{F}\lett{q}]$.
We now are left only with the two occurrences of $\sg{T}(\neg\lett{q})$ to treat, which we must do in two steps using the third equation of \allmodels. We write the final result at once:
$[\sg{T}\lett{p} , \sg{F}\lett{q} , \sg{F}\lett{p}] , [\sg{T}\lett{p} , \sg{F}\lett{q} , \sg{F}\lett{q}]$. 
Clearly the first set of literals is unsatisfiable which, by the way, we could have noticed some steps earlier, thereby obtaining a less expensive development. The second set is just 
$\{\sg{T}\lett{p} , \sg{F}\lett{q}\}$
\noindent and characterizes all the models of the original $\Gamma$.

There seem to be three inconveniences as to this execution. The first is that we have treated one and the same formula twice, and that on two different occasions. One readily realizes that the issue is avoidable if the use of the branching equation corresponding to a false conjunction is always subsequent to the use of every other (non-branching) equation formerly applicable.
The second inconvenience is that we have rewritten many a formula that was without change. And, finally, the execution is awkwardly traceable ---we have namely indicated the successive steps taken by means of narrative text interspersed in the successive rewritings. The latter is of importance when we consider executions by hand ---then a more formal and easily checkable notation would be most welcome by both students and teachers.
We shall consider these two remaining issues in the next two subsections, beginning with the latter about an easily traceable notation.


\nentry{Tree of lists.} The straightforward manner of making executions like the former traceable and easily verifiable is just to record the application of each rule, including mention to the formula used. We should therefore begin by naming the equations of the algorithm, say \ruleq{T\wedge}, \ruleq{F\wedge}, \ruleq{\neg} and \ruleq{l} in the order in which they are written above.
\noindent The procedure leads to the deployment of a tree structure whose nodes are lists of formulae as in the preceding section, and whose internal nodes (not leaves) are decorated by labels as explained presently: \begin{enumerate}  \setcounter{enumi}{-1}
\item To begin with, we have only one item, namely the original list of formula. This is of course a tree with only one terminal node (leaf).
\item At each step we choose a composite formula within a leaf (call this leaf $\mathcal{L}$) and apply the corresponding rule as already explained. As a result one or two new lists of formulae are obtained, which are linked to $\mathcal{L}$, becoming successors of $\mathcal{L}$ in the tree. At the same time we label $\mathcal{L}$ with the name of the equation and the formula used.
\end{enumerate}
The leaves of these trees coincide with the lists of formulae obtained by the procedure explained in the preceding paragraph ---we have only added a tree structure on top of them for tracing their computation. Therefore, the set of models of the root of the tree obtains as the union of the sets of models of the leaves. Formally, this much becomes clear after the consideration that the union of the sets of models of the leaves, and therefore the invariant just mentioned, are indeed preserved by each application of one equation as described above. Therefore the correctness of the
computation procedure using these trees follows by straightforward mathematical induction. The right formulation and proof of this result is left as exercise.

Notice that the preceding description amounts to inductively defining these trees as a family \ttabl{\Gamma} indexed by the finite sets $\Gamma$ in a manner such that the constructors stand in correspondence with the equations as named above, in the following manner: to internal nodes, constructors \ruleq{T\wedge}, \ruleq{F\wedge} and \ruleq{\neg} 
are associated, corresponding to the equation used in each case. The leaves are the as yet untreated nodes or those already formed by literals only. In either case we associate to the leaf the constructor \ruleq{l}. Unfortunately, we must skip a detailed explanation for reasons of space.

\nentry{Tree of formulae.}
The repetition of possibly large lists of formulae along the trees as introduced in the preceding section can be avoided, e.g. by employing the procedure described in \cite{Smullyan}. We describe these less expensive trees as follows. 
The general idea is to write at each node of the tree different from the root only the formulae originated by the use (decomposition) of another formula. The root of the tree will contain the originally given set (list) of formulae. With this information it is possible to compute the full trees of the preceding paragraph provided the used formulae are recorded at each step, i.e. at each node. Therefore, the correctness of the present method with improved trees will follow from the correctness of the prior method. 
Specifically, we define the improved trees as follows:\begin{enumerate}
\item Each node will have associated an explicit set $E$ 
 and an implicit set $\Gamma$ of formulae. $E$ is to be written down explicitly, whereas $\Gamma$ is to be computed when necessary.
\item For the root of the tree, both $E$ and $\Gamma$ coincide with the originally given set of formulae.
\item For the other nodes, $E$ will consist of one or two formulae.
\item All internal (i.e. non-leaf) nodes will also have associated one formula, to be called the one \textit{used} at the node.
\end{enumerate}
We now indicate how to extend the tree down from a leaf:
\begin{enumerate}
\item A formula in $\Gamma$ is chosen and written down at the node as its used formula.
\item Then one proceeds according to the form of the chosen formula:
\begin{enumerate}
\item In case it is of the form $\sg{T}\,(\alpha\wedge\beta)$ then the tree is extended with \emph{one} child node. For this new node, $E = \{\sg{T}\,\alpha , \sg{T}\,\beta\}$. 
\item In case it is of the form $\sg{F}\,(\alpha\wedge\beta)$ then the tree is extended with \emph{two} children nodes. One of them will have $E = \{\sg{F}\,\alpha\}$ 
, whereas the other one will have $E = \{\sg{F}\,\beta\}$.
\item Finally, in case the chosen formula is of the form $S\,(\neg\alpha)$ then the tree is extended with \emph{one} child node having $E = \{\op{S}\,\alpha\}$.
\end{enumerate}
For every case of newly created node, the set $\Gamma$ is computed as follows: If  $\Gamma_{0}$ is the implicit set of the parent node, then $\Gamma = (\Gamma_{0} - \sigma) \cup E$, where $-$ denotes deletion of a member in a set.
\end{enumerate}
Now, to each improved tree $t$ with a non-leaf root which has associated explicit set $E$ and implicit set $\Gamma$ of formulae, as well as used formula $\sigma$, a full tree of type \ttabl{\Gamma} can be associated, whose constructor is the one corresponding to the form of $\sigma$, i.e. \ruleq{T\wedge}, \ruleq{F\wedge} or \ruleq{\neg}, and its children trees are the ones (recursively) corresponding to the children trees of $t$.
If otherwise $t$ is just a leaf, then its corresponding full tree is $\ruleq{l}(\Gamma)$, where $\Gamma$ is the implicit set of formulae of the leaf in question.
This correspondence gives already a method for using the improved trees in order to compute all the models of any given set of formulae. Nevertheless, the following result makes such process easier:
The implicit set at each leaf is the union of the explicit sets at the branch ending up at the leaf in question, minus those formulae that have been used on that branch.
Thereby one can determine when a branch is \emph{completed}, which happens when the implicit set at the corresponding leaf is a set $\Gamma$ of literals. Further, then \lmods{\Gamma} is the corresponding set of models, and one can then compute the set of models of the whole tree (i.e. of the originally given set of formulae) by taking the union of the sets at each leaf, just as with the full trees.

\section{Conclusions}
We have put forward a presentation of classical propositional tableaux elaborated by application of some principles that are noteworthy in program design. Foremost among those principles is the one of \emph{separation of concerns}: We have namely started by deriving from a straightforward specification an algorithm given as a set of recursive equations for computing all models of a finite set of formulae. The correctness of the algorithm is brought about hand-in-hand with its derivation by means of a basic inductive argument whose cases are each solved by calculational reasoning yielding identities between sets of interpretations that need not the usual ``ping-pong'' (or direct-and-converse) argument.

Thereafter we discussed the employment of data structures, mainly with regard to a manual execution of the algorithm. A requirement of natural traceability and verification led us to the trees of sets or lists of formulae presented in \cite{Hintikka,Ben-Ari}, the correctness of which is immediate after their derivation as traces of the employment of the original equations. A further improvement avoids repetition of unmodified formulae giving rise to the trees presented in \cite{Smullyan}, whose correctness is in turn guaranteed by showing that they carry the same information as the former trees.

Smullyan's  classical presentation \cite{Smullyan} introduces instead the method as a proof procedure for establishing unsatisfiability of (finite) sets of (signed) formulae. The tableaux are given directly in the form of our improved trees of formulae. The proof of correctness is then as usual composed by two arguments, one of soundness and one of completeness, to the effect that unsatisfiable sets give rise to \textit{closed} tableaux, i.e. one in which every branch contains a contradiction and thus has no model. The proof of soundness is by a quite direct tree induction, whereas the proof of completeness involves showing that an open \textit{completed} branch, i.e. one in which every formula has been fully decomposed, is  a Hintikka set. Besides, Hintikka's lemma is proven, to the effect that every Hintikka set has a model.

In our experience, the use of the method as in the classical presentation leads students to the realisation that they either prove the given set of formulae inconsistent or can compute every counter-example (i.e. a sufficient characterisation thereof). Subsequently they tend to ask why we cannot establish such fact as a meta-theoretical result. Our presentation does precisely that ---and the correctness of the method as a proof procedure follows as immediate corollary. 
The idea of computing all models of the given set of formula has led us to give an abstract formulation of the procedure. 
We then treat as a separate matter the question of the concrete trace of the manual execution of the method. As we have been able to check, this treatment provides the students with improved command over the method, i.e. they exercise a more sound domain over what they are doing and also over the various possible notations or manners of justification they can give thereof.

It could be argued that Smullyan's presentation and proof is scalable to infinite sets of formulae and first-order-logic, and therefore ask about such feature regarding our presentation. Concerning infinite sets of formulae, the first thing to say is that  the validity of our equations is certainly not affected. Nevertheless, they cannot of course be interpreted anymore as an algorithm. Even if we assume as usual a principle of omniscience concerning the infinite sets, the method of choice of the formulae to be succesively decomposed by application of the equations is essential for getting the right result. But, as is the case also with the classical presentation, there exist method of orderly choice that guarantee (under the ominiscience principle) the computation of all models and thus the correctness of the method.
Generalisation to first-order logic, on the other hand, requires to abandon the idea of ``computing all models'', replacing it by e.g. ``determining whether the set of formulae is or not (un)satisfiable''.

We conclude that our presentation may contribute in a better way to the achievement of \textit{profficiency with understanding}, which is our main learning objective. Also it emphasizes design methodology, which we strive to do along and across the whole of  the program of studies.
It also could be argued that the method is tailored to just students of Computing Science or Software Engineering. We however believe that it can be taught also without much difficulty to Mathematics or Philosophy students and that the advantages we claim to obtain can also be appreciated in such cases. This, however, is yet to be checked out.

Finally, we should like to think of this work as one interpretation and case of the disclosing of the ``doing'' of Mathematics as advocated by Dijkstra \cite{Dijkstra}. We have tried to avoid all gaps of both mathematical and motivational nature. To our mind, this case is yet another sample of the unity of structure and method that mathematics and programming share\footnote{And that at a deeper level shows up in the propositions-as-types principle.}. Exploiting such unity should be fruitful for improving understanding and thus better helping learning.

\newpage
\thispagestyle{empty}
{\ }

\end{document}